\documentclass[useAMS,usenatbib]{mn2e}
\usepackage{epsfig,amsmath,amssymb,amsfonts,mathrsfs,latexsym,graphicx}
\newcommand{\gsim}{\, \raisebox{-0.8ex}{$\stackrel{\textstyle >}{\sim}$ }}

\begin{document}

\author[Watts \& Reddy]{Anna L. Watts$^1$ and Sanjay Reddy$^2$
\\ $^1$ Max Planck Institut f\"ur
  Astrophysik, Karl-Schwarzschild-Str.1, 85741 Garching, Germany;
  anna@mpa-garching.mpg.de \\
$^2$ Theoretical Division, Los Alamos National Laboratory, Los Alamos,
New Mexico 87545, USA; reddy@lanl.gov}

\title[Magnetar oscillations pose challenges for strange
stars]{Magnetar oscillations pose challenges for strange stars}

\maketitle

\begin{abstract}

Compact relativistic stars allow us to study the nature
of matter under extreme conditions, probing regions of parameter space
that are otherwise inaccessible. Nuclear theory in this regime is not
well constrained: one key issue is whether
neutron stars are in fact composed primarily of strange
quark matter.  Distinguishing the two possibilities, however, has been
difficult.  The recent detection of seismic vibrations in the aftermath of giant
flares from two
magnetars (highly magnetized compact stars) is a major breakthrough.
The oscillations excited seem likely to involve the stellar crust,
the properties of which differ dramatically for strange stars. We show
that the
resulting mode frequencies cannot be reconciled with the
observations for reasonable magnetar parameters. Ruling out strange
star models would place a
strong constraint on models of dense quark matter.
\end{abstract}


\maketitle

The properties of compact stars offer the best opportunity to
study the phase diagram of Quantum Chromodynamics (QCD) at extreme
densities, a region that is poorly constrained.   One hypothesis, based on the
possibility that strange quark matter (SQM) could be absolutely stable,  is
that compact stars are made almost 
entirely of deconfined SQM \citep{wit84}. In this
scenario, terrestrial matter made of nuclei is only metastable, albeit
with an essentially infinite lifetime because it would require
multiple weak interactions to convert normal matter to SQM
\citep{web06}. However, the extreme ambient 
conditions characteristic of compact stars can facilitate the
conversion of ordinary matter to SQM on short enough time-scales so
that some or all compact stars could be strange stars.  Distinguishing
strange stars from
neutron stars, however, has proved 
difficult. Signatures such as potentially smaller radii, and
a supposed inability
to glitch, are fraught with serious observational and theoretical  
uncertainty.   

New observations, however, could change this.  Certain
magnetars (compact stars with magnetic fields $\ge 10^{14}$ G), the 
Soft Gamma Repeaters (SGRs), exhibit gamma-ray 
flares powered by field decay \citep{dun92}.  Timing analysis of rare giant
flares, which have peak luminosities $10^{44}
- 10^{46}$ erg s$^{-1}$ and decaying tails lasting several minutes,
has recently revealed high frequency Quasi-Periodic Oscillations
(QPOs).  The 2004 giant flare of SGR
1806-20 shows QPOs at 18, 26, 30, 92, 150, 625 Hz and higher
\citep{isr05, wat06, str06}, whilst the 1998 giant flare from SGR 1900+14
has QPOs at 28, 53, 84 and 155 Hz \citep{str05}.   The most promising
model involves seismic vibrations, triggered by a 
starquake associated with the giant flare.  The lowest frequency 18, 26
Hz QPOs fit predictions for Alfv\'en modes of the core. For the higher
frequencies, 
attention has focused on toroidal shear modes of the crust (or their
global magneto-elastic equivalent).  These have frequencies that are, for neutron stars, a
good match to the observations \citep{mcd88,  
dun98, pir05, gla06, sam06}.  The 28-155 Hz QPOs  
would be $n=0$ modes (no radial node) with differing angular quantum
number $l$.  The 625 Hz QPO is consistent with being the $n=1$
first radial overtone.  

The likely dependence on crust properties is very
exciting. Originally, strange stars were expected to be devoid of a solid crust
and were 
characterized by an ultra-dense quark liquid extending up to the
surface \citep{hae86}. Such bare strange quark
stars could not account for torsional shear oscillations since there is
no solid region in the vicinity of the surface.  However, strange stars
can have solid crusts.  One 
possibility \citep{alc86} is that the strange star has a thin crust of normal
nuclear material extending down to neutron drip at density $\rho
\approx  4\times 10^{11}$ g cm$^{-3}$ (for a neutron star the crust extends
beyond neutron drip to $\rho \approx 10^{14}$ g 
cm$^{-3}$), suspended above  liquid SQM by an enormous
electric field.  A more recent model \citep{jai06} posits a crust in 
which nuggets of SQM are embedded in a uniform
electron background.   The strange star crusts have different shear speeds and
are thinner than neutron star crusts: both
factors affect shear mode frequencies, crust thickness $\Delta R$
being of particular importance to the radial overtones \citep{han80}.
Magnetar seismology may therefore offer a robust means of
distinguishing strange stars
from neutron stars.

\section{Model}
\label{mod}

Most previous studies computing torsional shear modes of the
stellar crust assume free slip over the fluid core.  But if the strong
magnetic field couples crust and core together then one 
should instead consider global magneto-elastic perturbations. Recent
work on the coupled problem, however, suggests that the vibrations
most likely to be observed have very similar frequencies to those
computed in the 
uncoupled problem \citep{gla06, lev07}\footnote{The recent calculation
by \citet{lev07} is particularly interesting, as it examines the
time-dependence of the interaction between the crust and the core in
response to an initial perturbation. The behaviour is complicated, with time-dependence that is not pure
oscillatory, but results in periodic amplification at the
natural crust frequencies.}.  The
natural crust frequencies are 
therefore physically relevant, so for simplicity we neglect coupling and 
apply zero traction boundary conditions \citep{car86}. 

Following \cite{pir05} we use a plane-parallel geometry with constant
gravitational acceleration $g = GM/R^2$ in the vertical direction $z$,
$M$ and $R$ being the mass 
and radius star.  The Newtonian equations of
hydrostatic equilibrium determine the crust density profile.
Using a slab rather than spherical geometry 
allows us to incorporate the magnetic field without difficulty; we
assume a constant 
field $\mathbf{B} = B\hat{z}$. For pure toroidal shear modes, which
are incompressible and have no vertical component of displacement, the
perturbation equations for the horizontal displacement $\xi$ reduce to

\begin{equation}
\frac{\left(\mu \xi^\prime \right)^\prime}{\rho}  + v_A^2 \xi^{\prime \prime} +
 \left[\omega^2\left(1
    + \frac{v_A^2}{c^2}\right) - \frac{(l^2+ l-2)\mu}{\rho R^2}\right] \xi
= 0
\end{equation}
where $\mu$ is the shear modulus and $v_A = B/(4\pi\rho)^{1/2}$ the
Alfv\'en speed.  Shear speed $v_s = (\mu/\rho)^{1/2}$.  We have
assumed a periodic time dependence 
$\exp(i\omega t)$, $\omega$ being the frequency.  We subsequently
correct for gravitational redshift to obtain the observed frequency.
Primes indicate 
derivatives with respect to $z$. We have used $\nabla^2_\perp \xi =
-[(l+2)(l-1)/R^2] \xi$ to mimic a spherical geometry, $l$ being the standard angular
quantum number.  The scaling with $l$ differs from that used by \cite{pir05}
but gives better agreement between slab and 
spherical models in the zero field limit \citep{mcd88,sam06}.   The
shear modulus $\mu$ \citep{str91} is  
 
\begin{equation}
\label{shear}
\mu = \frac{0.1194}{1 + 0.595(\Gamma_0/\Gamma)^2}\frac{n_i (Ze)^2}{a}
\end{equation}
Here $Z$ is the atomic number of the ions,
$n_i$ the density of ions and $a = (3/4\pi n_i)^{1/3}$ the average
inter-ion spacing. The parameter $\Gamma = (Ze)^2/ak_B 
T$, where $T$ is the temperature and  $\Gamma=\Gamma_0 = 173$ marks the point at
which the solid lattice melts to form an ocean
\citep{far93}, and we use this to determine the upper boundary of the
crust.  Crust temperature in the tail of a 
giant flare is not 
well constrained:  observations set a lower 
limit of $10^7$ K \citep{tie05}, but theory suggests that it could be
as high as $\sim 10^9$ K \citep{lyu02}, so we examine the range $T = 10^7 -
10^9$ K.  

We consider stellar models with $M = 1.2 - 2.5 M_\odot$ and $R = 8-15$
km, subject to causality constraints, to cover the full range of
possible strange star parameter space \citep{pag06}. Both thin nuclear and
nugget crust models are studied \citep{alc86, jai06}  We model magnetic fields in
the range $B = 10^{12} - 10^{15}$ G.  The strong field affects only
the very outermost low density layers of the crust \citep{har06}: we estimate the 
corrections to computed mode frequencies to be less than 1\%.  Energetic arguments show that global crust structure, including the
suspended crust model, is unaffected.  

For the thin nuclear crust there is some uncertainty in the
composition of neutron-rich nuclei, which will be constrained by
future rare isotope experiments.  We therefore survey a range of
equations of state \citep{hae94, rus06}.

The nugget crust is
composed of a lattice of strange quark nuggets embedded in a
background degenerate electron gas.  The electrons contribute to the
pressure, and the nuggets to the energy density. The density within
the crust
 is given by $\rho = x\epsilon_0$, where
$\epsilon_0=n_{\mathrm{quark}} \mu_q$ is the energy density of quark
matter inside nuggets and depends on
the density $n_\mathrm{quark} \approx 1$ fm$^{-3}$ and chemical
potential $\mu_q \approx$ 300 MeV at which stable quark matter
vanishes.  $x$ is the volume fraction occupied by nuggets and is given
by   
\begin{equation}
x= \frac{\mu_e^3}{3\pi^2(n_Q - \chi_Q \mu_e)} \,,
\end{equation}
where $\mu_e$ is the electron chemical potential,  $n_Q $ is the
electric charge density of the quark nugget and  $\chi_Q$ is its
charge susceptibility \citep{jai06}. The quark matter parameters are
poorly known and can only be determined within the context of specific
models. In the Bag model, $n_Q=m_s^2 \mu_q/2\pi^2$ and
$\chi_Q=\mu_q^2/\pi^2$. Further, requiring SQM to be absolutely stable
and simultaneously requiring normal nuclei to be metastable restricts
$\mu_q$ to a narrow range centered around $\mu_q \sim 300 $ MeV. Thus
within the Bag model, the remaining uncertainty is parametrized
through the effective strange quark mass $m_s$ which we expect to be
in the range $150-250$ MeV.    

The spherical nugget phase occupies most of the
crust; we will neglect the small region at the base of the
crust containing the pasta
phase (see below).
In computing shear modulus (Eqn. \ref{shear}) the quantity \begin{equation}
Z = \frac{4\pi}{3} f R_n^3  (n_Q - \chi_Q \mu_e) 
\end{equation}
is the charge of the nugget where $R_n = y~\lambda_d$ is the typical nugget size, $\lambda_d = 1/(4\pi\alpha \chi_Q)^{1/2}$ being the Debye screening length 
and $\alpha$ the fine structure constant.  The factor $f$ is a
correction due to screening inside nuggets, $f \approx 3(y - \tanh
y)/y^3$, where $y \approx 1.6$ \citep{alf06} gives $f \approx 0.5$ for
the larger droplets. The quantity $a = R_n/x^{1/3}$ is the average
inter-nugget distance, and $n_i = 3/(4\pi a^3)$ the density of
nuggets. 

The variation of $\mu_e$ with depth is given by integrating the
equations of hydrostatic equilibrium.  Using the limits $x=0$ and
$x=1$, \cite{jai06} followed this procedure to estimate $\Delta R$.
In reality the region with $x \gsim 0.5$ that contains the pasta phase
occupies a tiny region
with thickness $\lesssim 1$m.  Further, with increasing $x$ the free energy gain of the
heterogeneous state becomes negligible and even a small surface
tension can disfavor the large $x$ region. For these reasons we set
$x\simeq 0.5$ at the base of the crust. The value of $x$ at the top of
the crust is set by the melting condition: for
the parameters examined, it lies in the range $x \sim
10^{-12} - 10^{-4}$.     

\begin{figure}
\begin{center}
\includegraphics[width=8cm, height=6cm, clip]{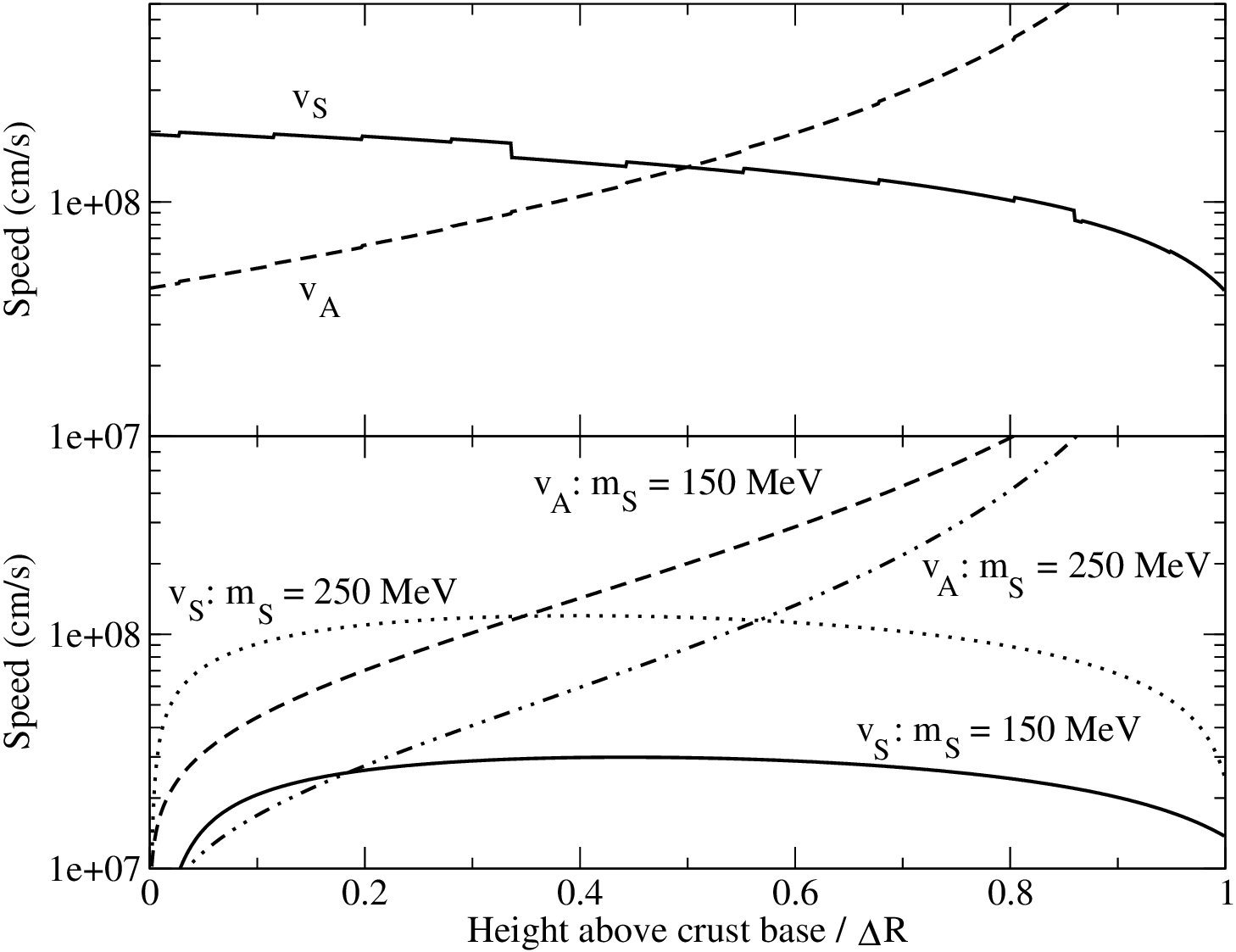}
\end{center}
\caption{Shear and Alfv\'en velocities ($v_s$, $v_A$) in the crust for
  stars with  $M = 1.4 M_\odot$, $R = 12$ km, $T = 10^8$ K and $B =
  10^{14}$ G.  Top:  Thin nuclear crust (crust thickness $\Delta R/R$
  = 5.4\%).  Bottom:  Crust with nuggets, for $m_s = 150$ MeV
  ($\Delta R/R$ = 0.7\%) and $m_s = 250$ MeV ($\Delta R/R$ = 6.1\%).}  
\label{f1}
\end{figure}

Figure \ref{f1} shows the variation of $v_s$ and $v_A$ 
with depth for example crust models (compare to Fig. 1 of
\cite{pir05}). The shear speed in the nugget crust is smaller than in
the  thin-nuclear crust. This can be understood by noting that at
constant pressure $v_s \sim \sqrt{Z^{5/3}/A}$ where $A$ denotes the
baryon number. Further, and unlike in the nuclear case, both $Z$ and
$Z/A$ of the nuggets decrease rapidly with depth.

\section{Results}
\label{res}

Our results are good agreement with analytic estimates \citep{mcd88,
  dun98, pir05, sam06}.  Overall, frequencies scale with $v_s$.  The
  $n=0$ mode frequencies are almost independent of $\Delta R$,
  and varying $B$ and $T$ changes frequency by no more than a few Hz.
  The frequencies scale as $[(l+2)(l-1)]^{1/2}$, making it difficult to fit a mode sequence with an 18 or 26 Hz fundamental.
Following previous studies we assume that these represent
  global Alfv\'en modes, and search for a fundamental at $\approx 30$
  Hz.   The radial overtones vary little with $l$, but
depend strongly on $\Delta R$, frequencies tending to increase as the crust
  thins. Crust thickness is set in part by compactness. Increasing
  temperature also tends to push overtone frequencies up as 
  the outer layers of the crust melt.  The overtones
also depend strongly on $B$, since $v_A$ exceeds $v_s$ at magnetar field
strengths.    

For the thin nuclear crust models the residual uncertainty in
the equation of state leads to variations of at most a few percent,
and in what follows we quote results based on
\cite{hae94}. For the parameter space studied the frequency of the $n=0, l=2$ mode
lies in the range 26 - 54 Hz.  To obtain a fundamental $\le 30$ Hz
requires high mass ($M \ge 2.4 
M_\odot$ for all radii, or $M \ge 2.2 M_\odot$ and $R \ge 14$ km),
pushing the limits of strange star parameter space \citep{pag06}.   

\begin{figure}
\begin{center}
\includegraphics[width=8cm,height=5cm, clip]{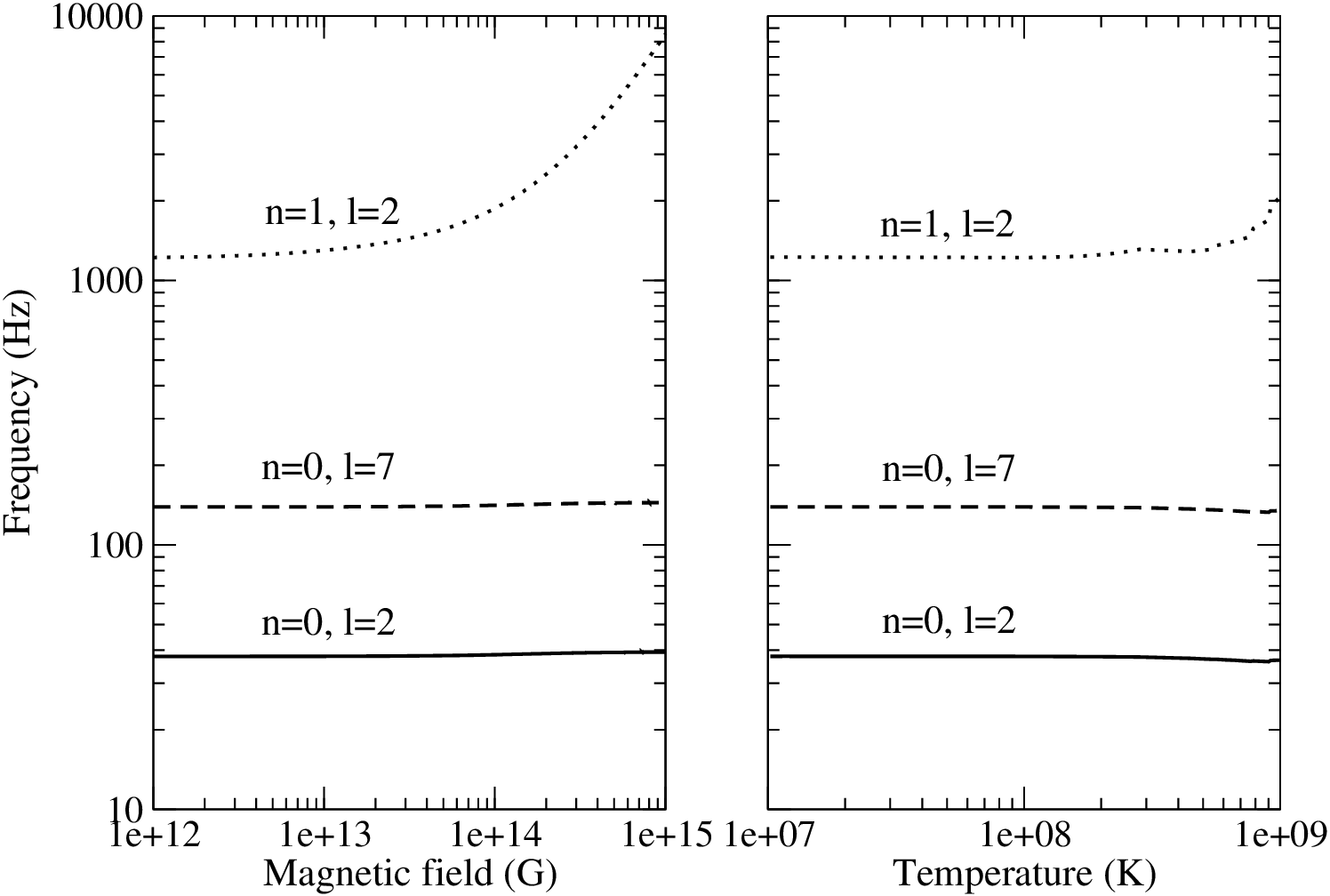}
\end{center}
\caption{Thin nuclear crust mode frequencies (for a star with $M =
  1.4M_\odot$, $R = 12$ km).  Left:  effect of varying magnetic field,
  for fixed $T = 10^8$ K.  Right:  effect of varying temperature, for
  fixed $B = 10^{12}$ G. Crust thickness $\Delta R/R$ falls from 5.4\%
  at $T = 10^8$ K to 2.1\% at $T = 10^9$ K.}  
\label{f2}
\end{figure}

The overtone
frequencies are all high.  Even at $B= 10^{12}$ G, the lowest 
frequency is $\approx 725$ Hz, for the model with the thickest crust ($M
= 1.2 M_\odot$, $R = 15$ km, $\Delta R/R$ = 8\%).   For $B >
10^{14}$ G, the lower limit on overtone frequency is higher
still, at $\approx 1100$ Hz. There is no
model that would allow an overtone at 625 Hz.  Figure \ref{f2}
illustrates the effects of varying $B$ and  
$T$ on the frequencies of both the $n=0$ and $n=1$ modes. 

For the nugget crust models there is additional uncertainty in the
value of $m_s$.  Increasing $m_s$ increases shear speed, and for the
$n=0$ modes frequency scales directly with $m_s$. However, in general
$v_s$ is lower than for the nuclear crust models.  For the parameter
space studied the frequency of the $n=0, l=2$ fundamental lies in the
range $\approx 1 - 21$ Hz.  There is no model that would permit a
fundamental in the range 28-30 Hz. 

\begin{figure}
\begin{center}
\includegraphics[width=8cm,height=5cm, clip]{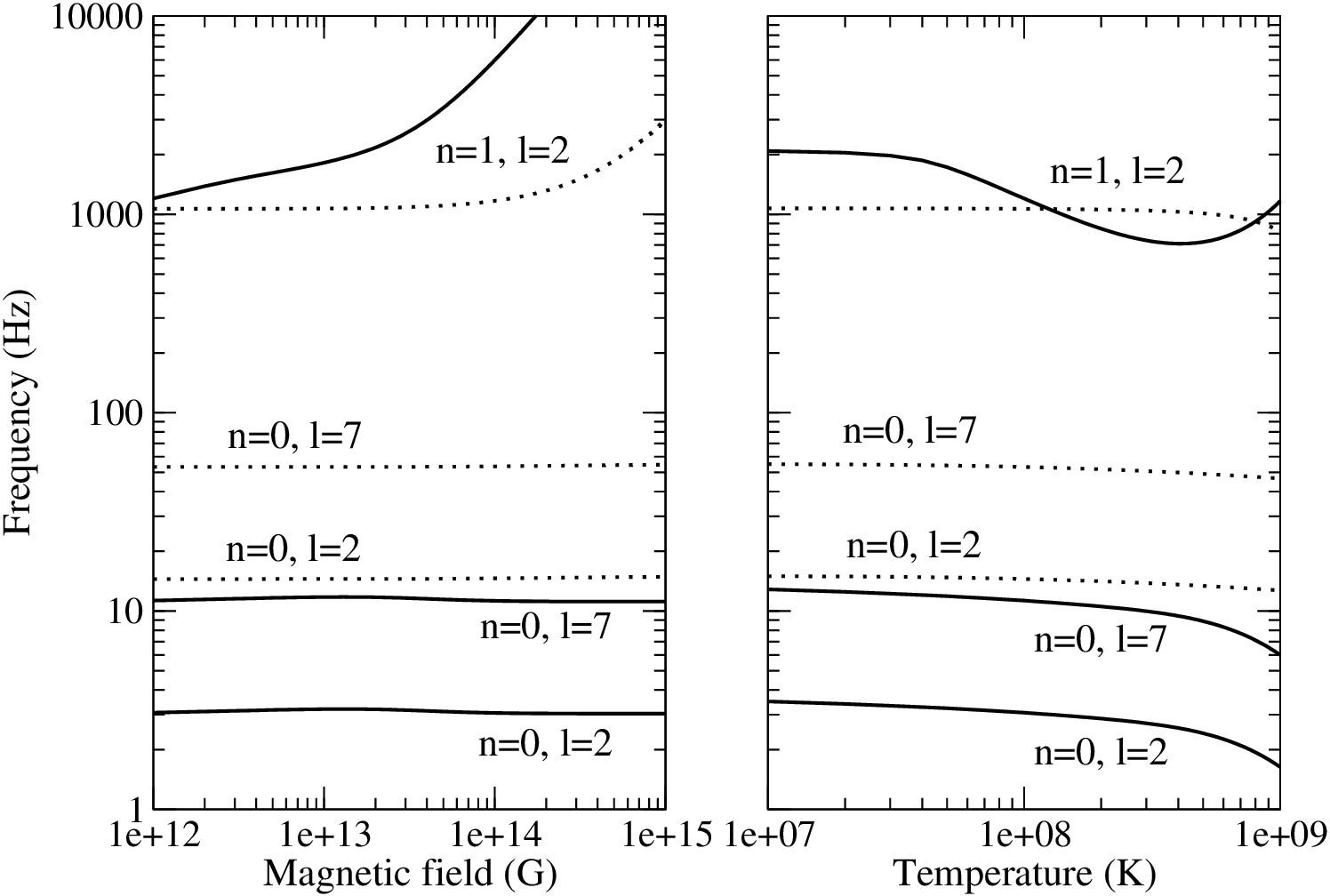}
\end{center}
\caption{Nugget crust mode frequencies (for a star with $M =
  1.4M_\odot$, $R = 12$ km)  for strange quark mass $m_s = 150$ MeV
  (solid) and $m_s = 250$ MeV (dotted). Left:  effect of varying
  magnetic field, for fixed $T = 10^8$ K.  Right:  effect of varying
  temperature, for fixed $B = 10^{12}$ G.  The minimum in frequency
  for the overtone occurs at higher temperatures for $m_s = 250$ MeV.
}   
\label{f3}
\end{figure}

The range of possible frequencies for the $n=1$ overtones is extremely
large. Crust thickness increases dramatically with $m_s$, and this effect can
in fact compensate for the change in $v_s$.  For the highest 
value of $m_s$, $\Delta R/R$ is comparable to or thicker than the thin
nuclear crust; for the lowest values, however, it can be an order of
magnitude thinner.  Temperature dependence is complex.  As $T$
increases from $10^7$ K frequency drops as $v_s$ drops, the effect
being more pronounced than for the nuclear crust.  However at high
enough $T$ crust thinning accelerates dramatically.  This can
offset the effects on $v_s$ and cause frequency to rise again. At $B =
10^{12}$ G, frequencies for the $n=1$ overtone can be as low as
$\approx 260$ Hz.  Frequencies do however rise as $B$ increases, and
by $B=10^{14}$ G only a few very extreme models permit an overtone
frequency as low as 625 Hz ($M = 1.2 M_\odot$, $R = 15$ km, $m_s
\approx 250$ MeV, $T \ge 9 \times 10^8$ K). Figure \ref{f3} illustrates the
effects of varying $B$, $T$ and $m_s$  
on the $n=0$ and $n=1$ modes.  

\section{Discussion}
\label{disc}

For neutron star models torsional shear modes, or their global
magneto-elastic equivalents, are a good fit for the observations.  For
strange stars, the situation is much more difficult.   A thin nuclear crust model
can give a 
fundamental in the right range if stellar mass is large, but
the overtone frequencies are far 
too high.  The nugget crust model permits a wider range of
frequencies due to the uncertainty in the strange quark mass $m_s$.
The lowest order $n=0$ modes could explain some of the QPOs in the 
range 18-150 Hz, but because the fundamental frequencies are so low it
is difficult to fit a mode sequence, given the expected scaling
with $l$ \citep{sam06}. For the radial overtones there are
regions of parameter space at high temperature with modes at the right
frequency, but only for magnetic fields at least an order of magnitude lower
than those inferred for magnetars.  At magnetar field strengths only
models at the very limits of parameter space permit an overtone frequency
as low as 625 Hz.  The other
constraint on the nugget crust is the sensitivity of the overtones to
temperature fluctuations (greater for this model than for the thin 
nuclear crust or  
neutron star crusts).  The observations indicate that the 625 Hz QPO has high
coherence and lasts for several hundred seconds, a period during which
temperature could vary substantially \citep{wat06, str06}.
   
We conclude that the frequencies of toroidal shear modes in strange star crusts
have serious difficulty explaining the QPO frequencies observed during magnetar
hyperflares.  If the results of \cite{gla06} and \citet{lev07} hold
true, this
conclusion will not change greatly when coupling to core is included
in the calculation.  However, these results must now be verified using more
sophisticated models that include a realistic stellar geometry,
field configuration, and
general relativistic corrections.   There may also be other types of
modes in the right 
frequency range \citep{chu06},  although these should be harder to
excite and detect.  We note that there are alternative non-seismic
models for the QPOs, but that these models have serious difficulties (for a discussion see
\cite{wat06b}). 

 The clear distinction between the theoretical predictions for neutron star and
 strange star crust models is extremely promising.  It offers a
 robust, largely model independent means of distinguishing strange
 stars from neutron stars,
 something that has in the past been lacking.  This type of study
 should lead to rapid progress in constraining the 
 equation of state of compact stars.   Ruling out the strange star hypothesis
 would directly impact the phase diagram 
of QCD at finite chemical potential.  It would offer a strong constraint on
models of dense  
quark matter, indicating that the deconfinement transition is not 
significantly lowered by the dynamics of the strange quarks. One
important implication is that multiply strange hadronic states such  
as the H-dibaryon in the terrestrial context are less likely. In addition, 
constraints on the models at finite chemical potential should be relevant 
for finite temperature extensions of these models which have been employed 
to describe heavy-ion experiments such as the Relativistic Heavy-Ion
Collider (RHIC). Perhaps most importantly, 
ruling out strange stars directly from observations is the only way to 
ascertain that terrestrial matter is stable.

\section{Acknowledgments}

We thank J.Lattimer, N.Rea, A.Piro, C.Thompson,
P.Woods, H.Spruit, L.Samuelsson, T.Strohmayer and J.Schaffner-Bielich for comments.  ALW was supported by the EU FP5 Research Training Network `Gamma-Ray Bursts:  An Enigma and a Tool', and SR by the DoE under Contract No. W-7405-ENG-36.

\end{document}